\documentclass{ws-procs975x65}

\newcommand{\req}[1]{Eq.\,(\ref{#1})}
\newcommand{\beqn}{\begin{equation}}
\newcommand{\eeqn}{\end{equation}}

\begin{document}

\title{Critical Acceleration and Quantum Vacuum}

\author{Johann Rafelski$^*$ and Lance Labun}

\address{Department of Physics, The University of Arizona,\\
Tucson, AZ 85721, US}


\begin{abstract}
\vskip -0.2cm Little is known about the physics frontier of strong acceleration; both classical and quantum physics need further development in order to be able to address this newly accessible area of physics. In this lecture we  discuss what strong acceleration means and possible experiments  using electron-laser collisions and, data available from ultra-relativistic heavy ion collisions. We review the foundations of the current understanding of charged particle  dynamics in presence of critical forces and discuss the radiation reaction inconsistency in electromagnetic theory and the apparent relation with quantum physics and strong field particle production phenomena.  The role of the quantum vacuum as an inertial reference frame is emphasized, as well as the absence of such a `Machian' reference frame in the conventional classical limit of quantum field theory.
\end{abstract}

\keywords{acceleration, quantum vacuum, electron-laser collisions, Mach's principle}
\bodymatter

\section{Introduction}
Special relativity guarantees that all inertial frames of reference are equivalent. However, we do not know which body is inertial and which is accelerated since there is no apparent connection of the classical microscopic laws of physics to a global class of inertial reference frames, which provides  the required definition of inertial motion.  On the other hand, when formulating the quantum field theory we must introduce  in addition to  the quantum action also the ground state,  the quantum vacuum. This vacuum state provides within the theory the reference to a general inertial frame. In that sense the quantum theory seems to be considerably closer to `knowing'  `which is the accelerated frame'. However,   this information is lost in the present day procedure for passing to the  classical limit. 

A related challenge in understanding physics at high acceleration is that there is no limit to the strength of force and thus no limit to the acceleration that the inertia of a body can be subject to. To realize acceleration within  e.g. electromagnetic (EM) theory, electric and magnetic fields are used. As the strength of the applied electric field increases, so does acceleration imparted to a particle. However, within quantum electrodynamics (QED) when the field strength is too large, a rapid conversion of field energy into pairs ensues, weakening the field and establishing an effective limit to acceleration strength. 

In this report we introduce these ideas in greater detail, discuss open questions in the current theoretical framework, and propose methods to explore experimentally new physics emerging. We  discuss how the  concept of critical acceleration unites different disciplines of physics in which one speaks of critical field strength. Critical acceleration can be today achieved both in ultra-intense laser pulse collisions with relativistic electrons and in ultra relativistic heavy ion collisions at RHIC and at LHC.  We survey the classical theory of electromagnetism and discuss the shortcomings related to the problem of radiation reaction and the related problem of electromagnetic mass contained in the field.  We discuss in depth the important role of the quantum vacuum as the inertial reference frame.
  
\section{Critical Acceleration}
\subsection{Definition}
Critical acceleration of unity in natural units
\beqn\label{ac}
a_c=1\equiv \frac{Mc^3}{\hbar} \to   2.331\:10^{29} \frac{\rm m}{{\rm s}^2} {\rm \ \ for\ } M=m_e
\eeqn
contains implicitly the inertial mass of the particle being accelerated. Therefore one may introduce   critical  specific  acceleration
\beqn\label{aleph}
\aleph_c=\frac{a_c}{Mc^2}=\frac{c}{\hbar}.
\eeqn 

Both $a_c$ and $\aleph_c$ are constructed employing the same fundamental constants we see in defining Planck mass or length.  In addition, the  gravitational constant $G_N$, which establishes the relation of mass-energy density and geometry, is required in defining the Planck length and mass.  In general relativity, $G_N$ can be expected to connect with acceleration. However, consider a Newtonian force acting at the Planck length
\beqn\label{alephN}
\aleph_c^{N}=\frac{G_N}{L_p^2}=\frac{G_N}{\hbar G_N/c}=\aleph_c
\eeqn 
The critical specific acceleration $\aleph_c$ arises from a Newtonian force between two Planck masses separated by one Planck length. Notably, the gravitational constant $G_N$ cancels. Even though the value of critical acceleration is gigantic, the absence of $G_N$ opens up the possibility of present day experiments at the `Planck' acceleration scale.

By virtue of the equivalence principle, we probe particles subject to Planck-scale force in non-gravitational interactions whenever $a_c, \aleph_c$ is achieved. In order to achieve critical accelerations we need  strong `critical' fields  made possible by the formation of extended material objects, not present in Einstein's general relativity (GR) theory of dynamics of point particles. It is the quantum theory combined with gauge interactions which creates a resistance to free fall. Free fall would be otherwise the natural state of any GR particle system. Free fall of all particles is evidently the point of view taken by Einstein, e.g. in his study of GR solutions   for the case of radial motion of a dust of massive particles~\cite{Einstein:1939}.

\subsection{Experimental Methods}
\subsubsection{Relativistic heavy ion collisions}
Critical acceleration can be  attained in many  high energy hadronic and heavy ion collisions. The phenomenon of interest in this context is the rapid stopping of a fraction of matter in the projectile and target hadron in the CM frame of reference. The acceleration required to stop a constituent of colliding hadrons is estimated from the rapidity shift $a\simeq \Delta y/M_i\Delta\tau$ with $M_i\simeq M_N/3\simeq 310\:{\rm MeV}$.  We consider an example collision of ultra-relativistic heavy ions.  With $\Delta y=2.9$ at the SPS or $\Delta y=5.4$ at RHIC, $\Delta\tau$ must be less than 1.8 fm/c at SPS or 3.4 fm/c in order to have $a>a_c$. 

While there is no direct experimental evidence that these limits are satisfied, the global evidence from many related experimental efforts is $\Delta\tau<1$\,fm, yielding the preliminary conclusion that critical acceleration phenomena  are probed in these interactions.  The observation of an excess of soft photons in such experiments~\cite{Wong:2010gf} may present already a signal of new physics.

\subsubsection{Electrons in strong fields}
Identification of novel physics  phenomena  in the context of strong interactions is complicated by the many particles created and the different energy scales involved.  We can achieve cleaner experimental conditions exploring the behavior of  an electron in electromagnetic fields.

For the electron,   $a_c$  is achieved subjecting  an electron to an electrical  field which has the  Schwinger critical field strength
\beqn\label{Ec}
E_c=\frac{m_e^2c^3}{e\hbar}=1.323 \times 10^{18} {\rm V/m}.
\eeqn
This field strength has long been recognized as critical because an electric field of this strength is expected to decay quickly into copious electron-positrons pairs~\cite{Labun:2008re}.

\subsubsection{Relativistic laser pulses}
Instead of seeking to create in lab an electric field of magnitude \req{Ec}, we can boost the field and hence the acceleration of an electron by setting up a high energy electron-laser collision.  The demonstration experiment of this type was undertaken at SLAC in the late 1990s.  The electron energy was $E=\gamma m_ec^2=46.6$ GeV. A laser of greatest intensity at the time was employed~\cite{Burke:1997ew,Bamber:1999zt} with $a_0=0.4$, where the dimensionless  amplitude  is  defined by
\beqn\label{azero}
\vec A={\cal R}e\left(\vec A_0e^{i(\vec k\cdot \vec r-\omega t)}\right), \quad
a_0\equiv |\vec A_0|\frac{e}{mc^2}.
\eeqn  
Together these values of $\gamma, a_0$ imply that a peak acceleration of $|\dot u^{\alpha}|=0.073\:[m_e]$ was achieved. In these conditions the experiment recorded the effects of nonlinear Compton scattering and electron-positron pairs created by the Breit-Wheeler process.  

Today high-intensity laser systems are available with $a_0\gtrsim 50$, that is $(100)^2$ greater intensity.  Further, in addition to SLAC, there is CEBAF with a 12 GeV electron beam, radiation-shielded experimental hall and a laser acceleration team already associated with the facility.  Renewing this experimental concept provides in our opinion an  immediate access to beyond critical acceleration physics.

\section{Electromagnetism and Radiation Reaction}
\subsection{Independence of Fields and Particles}
The physics issues arising at high acceleration can be well illustrated considering the dynamics of charged particles interacting with a strong electromagnetic field. This situation is described by the Maxwell-Lorentz  action
\beqn\label{EMaction}
{\cal I} = -\frac{1}{4}\! \int \! d^4x \:F^2 
+\int \! d^4x \!\sum_i \! q_i\!\! \int_{{\rm path}_i} \!\!\!\!\!\!\!d\tau\,  u_i \cdot A(x) \delta^4(x-s_i(\tau_i))   
+ \sum_i\frac {m_{\rm I}^i c}{2} \int_{{\rm path}_i}\!\!\!\!\!\! d\tau \,(u_i^2-1).
\eeqn
There are three separate components, but only two at a time are involved in the generation of, respectively,  field dynamics (Maxwell equations) and the particle dynamics (Lorentz force).  

The Maxwell field equations are obtained by varying the first two components with respect to $A^{\alpha}$, where $F^{\beta\alpha}=\partial^\beta A^\alpha- \partial^\alpha A^\beta$,
\beqn\label{Maxeq}
\partial_\beta F^{\beta\alpha} = j^\alpha\:, \qquad  \partial_\beta F^{*\,\beta\alpha}=0\to \quad F^{\beta\alpha}(x),
\eeqn
and the source of the field is due to all charged particles
\beqn \label{MLSource}
  j^\alpha(x)=\sum_i\int d\tau_i\: u_i^\alpha q_i\delta^4(x-s_i(\tau_i)).
\eeqn
As indicated in \req{Maxeq} the solution for a given source is the field $ F^{\beta\alpha}$.

The gauge invariance of the first term in the action is assured by the fact that it depends on the fields only. However the middle term which relates particle inertia to the field is not manifestly gauge invariant.  Inserting the gauge potential $A\to \partial \Lambda$ we find that this term is a total differential: 
\beqn\label{GT}
\int \! d^4x \!\sum_i \! q_i\!\! \int_{{\rm path}_i} \!\!\!\!\!\!\!d\tau\,   \frac{ds_i}{d\tau} \cdot\frac{\partial \Lambda}{\partial x} \delta^4(x-s_i(\tau_i))
=- \!\int \! d^4x \!\sum_i \! q_i\!\! \int_{{\rm path}_i} \!\!\!\!\!\!\!d\tau\, \frac{d}{d\tau} \delta^4(x-s_i(\tau_i))
\eeqn
For each particle, the initial and final value of $\tau_i$ is chosen to correspond to the respective instances that the particle crosses the hypersurface space-time volume integrated over. Then the right hand side is a complicated way to say that the sum of charges entering the integration domain is the same as the sum of charges exiting from it. Therefore the condition that the gauge potential does not contribute to the action is that charge is conserved, that is in differential form,  $\partial\cdot j=0$.  Evaluating this by means of \req{MLSource} we of course recover the condition \req{GT}.

The Lorentz force is obtained by varying the  two components on right in \req{EMaction} with respect to the particle world line. However, to preserve the gauge invariance of the result we must not allow a variation at the surface of the domain, so as to guarantee that the result is compatible with both gauge invariance and charge conservation. One finds
\beqn \label{Lorentzeq}
m_{\rm I}\frac{du^{\alpha}}{d\tau} = -qF^{\alpha\beta}u_{\beta};
\quad  \frac{d s^\alpha}{d\tau}\equiv u^\alpha(\tau) \to s^\alpha(\tau),
\eeqn
As indicated, for a given field we can obtain the path of each particle. 

The above is a summary of book material. Yet there is an obvious challenge not all books discuss.  The question is,  are the three terms in the Maxwell-Lorentz action \req{EMaction} for critical acceleration consistent with each other and leading to consistent  dynamics of particles and fields? This question does not have an obvious answer: the action \req{EMaction} is relativistically invariant and leads to gauge invariant dynamics by intricate construction. However, otherwise it appears as an ad hoc composition of several terms.   In fact we recapitulated here the  derivation of particle and field dynamics to emphasize how they arise in a  separate and distinct manner from the action and variational principle framework but involve quite distinct objects of variation, fields and paths.

\subsection{Radiation Reaction}
A simple example addressing this consistency is the motion of a charged particle in a magnetic field according to Lorentz equation.  While the direction of motion changes, the particle energy remains constant.  However, the acceleration that is required to change the direction of motion causes radiation, and emission of radiation removes energy from particle motion placing it in the field.  The energy loss is a `small' effect for small fields and accelerations.  When the magnitude of the acceleration approaches the critical value \req{aleph}, the dynamics of charged particles are decisively influenced by the radiation field.  This is called radiation reaction: to describe how a particle moves we must account for its generated radiation field in addition to  the applied strong magnetic field.

The dynamical equations  Eqs.\:\eqref{Maxeq} \&  \eqref{Lorentzeq} can be `improved' to account for radiation reaction.  The idea pursued by Abraham and Lorentz  is to solve the Maxwell equations  using Green's functions, obtain radiation field, and incorporate the emitted radiation field as an additional force in the Lorentz equation.  This program leads to the Lorentz-Abraham-Dirac (LAD) equation written in this   form by Dirac~\cite{Dirac:1938nz}
\beqn\label{LADeqn}
\frac{(m_{\rm I}+m_{\rm EM})du^\alpha}{d\tau} = u_\beta q(F_{\rm external}^{\beta\alpha}+F^{\beta\alpha}_{\mathrm{rad}}),\qquad
F^{\beta\alpha}_{\mathrm{rad}}
= \frac{2q}{3}(\ddot u^{\beta}u^{\alpha}-\ddot u^{\alpha}u^{\beta}) 
\eeqn
Here the $F_{\rm external}^{\beta\alpha}$ is the field generated by all other charged particles, i.e. a prescribed field in which the particle considered moves. We see two effects, the appearance of $F^{\beta\alpha}_{\mathrm{rad}}$ which is the radiation field generated by the motion of the particle considered, and $m_{\rm EM}$ which is the classical electromagnetic energy content of the field a charged particle generates.   Note that at critical acceleration, the radiation reaction correction $F^{\beta\alpha}_{\mathrm{rad}}$ has the same order of magnitude as the Lorentz force, and thus in principle the iterative feed-back  used to obtain  radiation reaction in LAD from \req{LADeqn} breaks down.

\subsection{Problems with LAD}
The LAD equation of motion presents two foundational challenges.  First, a charged particle mass acquires an electromagnetic component, which arises from the energy content  of particle's field.  Second, there are solutions of LAD that violate causality. \\
{\bf 1)}  The electromagnetic mass $m_{\rm EM}$, appearing next to the inertial mass $m_{\rm I}$, is divergent in  Maxwell's electromagnetism.  This divergence can be regulated by establishing a limiting field strength within a modified framework such as in Born-Infeld~\cite{Born:1934gh} (BI) or another nonlinear theory of electromagnetism.  In practical terms we modify the first term in \req{EMaction}.

The particular attractiveness of the BI theory is the elegant format of the action, which addresses the   possible presence of curved space time:
\begin{align}\notag
{\cal I}_F = &- \! \int \! d^4x \sqrt{-\det\left(g_{\mu\nu}\right)}\:\frac{F^2}{4} 
\to -\! \int \! d^4x \left(\sqrt{-\det\left(G_{\mu\nu}\right)}-\sqrt{-\det\left(g_{\mu\nu}\right)}\right) E_{\rm BI}^2 \\
 &G_{\mu\nu}=g_{\mu\nu} +\frac{F_{\mu\nu}}{E_{\rm BI}},\quad
 \det\left(G_{\mu\nu}\right)=-1+\frac{E^2-B^2}{E_{\rm BI}^2}+\frac{(E\cdot B)^2}{E_{\rm BI}^4}
\label{BornInf}\end{align}
We evaluated the determinant of $G_{\mu\nu}$ in flat space. Like in any theory of nonlinear electromagnetism, the Maxwell equations apply to displacement fields $D, H$, while the Lorentz force depends on the $E, B$ fields.  The nonlinear relation between  $D, H$  and $E, B$ within BI theory imposes a limit on the electric and magnetic field strength and thus an upper limit on the force and acceleration. 

The problem with the BI approach is that precision tests show that the linear Maxwell theory still applies for very large `nuclear' field strengths. Therefore even if we were to assume that all of the electron's mass resides in $m_{\rm EM}$,  the predicted electron mass would have to be much larger than observed~\cite{Rafelski:1973fm}. That is, the limiting field $E_{\rm BI}$, required to `explain' electron mass as being electromagnetic is too small to be consistent with other experimental evidence. \\
{\bf 2)} The appearance of a third derivative $\ddot{u}^\alpha,\ u^\alpha=\dot x^\alpha$  in \req{LADeqn} requires  assumption of an additional boundary condition to arrive at a unique solution describing the motion of a particle. Only a boundary condition in the (infinite) future can eliminate solutions exhibiting self-accelerating motion, that is `run-away' solutions. Such a constraint is in conflict with the principle of causality.

There is no known natural remedy to the LAD problems in a systematic ab initio process that for example involves choosing a new action for the charged particle system.  As noted, LAD itself is not fully consistent as it was derived assuming that the effect of radiation reaction is perturbative.  The effort of Born and Infeld to modify the field action in the end did not cure either of the two problems. Modifying the field-particle coupling is very difficult seeing the subtle implementation of gauge invariance, and we know of no modification of the inertial term consistent with the Lorentz  invariance.  A commonly heard point of view is to negate the existence of a problem arguing that classical dynamics are superseded by quantum physics. However, strong acceleration relates particle dynamics to gravity, which is a classical theory. {\em The only way that quantum physics could help is to generate a theoretical classical limit that differs from our expectations. We will return to this point below.}

\subsection{Landau-Lifshitz Equation}
The second  difficulty with the LAD equation of motion \req{LADeqn} motivated many ad hoc repair efforts in the intervening years~\cite{Brasov}.  Among them, the approach of Landau and Lifshitz (LL)~\cite{LandauLifshitz} has attracted much study because it has no conceptual problems, is  semi-analytically soluble~\cite{Rivera,Rajeev:2008sw,DiPiazza,Hadad:2010mt}, and incorporates the Thomson (classical) limit of the Compton scattering process~\cite{Padmanabhan97}. The LL form of radiation reaction originates in   perturbative expansion in the acceleration, with the problematic third derivative replaced according to
\beqn\label{LLmod}
\ddot{u}^{\mu} \to \frac{d}{d\tau} \left( -\frac{e}{m} F^{\mu\nu} u_{\nu} \right).
\eeqn
The resulting equation of motion ($m$ is the sum of inertial and electromagnetic mass)
\beqn\label{LLeom}
m  \, \dot{u}^{\mu}=-\frac{e}{c} F^{\mu\nu} u_{\nu} -\frac{2e^3}{3m} \, 
\left( \partial_{\eta} F^{\mu\nu}u_{\nu}u^{\eta}  -\frac{e}{m}F^{\mu\nu} F_{\nu}^{\eta}u_{\eta} \right)
+ \frac{2e^4}{3m^2} F^{\eta \nu} F_{\eta\delta} u_{\nu} u^{\delta} u^{\mu}
\eeqn
is equivalent to LAD only for weak accelerations, a point stated not sufficiently clearly by Landau and Lifshitz~\cite{LandauLifshitz}. It is important to recognize that the LL equation \req{LLeom} implies the field-particle interaction is altered.  However, an appropriate fundamental action has not been found.  It must therefore be studied at the level of the equation of motion.  

We have studied the motion generated by \req{LLeom} for a laser-electron collision~\cite{Hadad:2010mt}.  In a consistent solution to the coupled LL dynamics, only after radiation loss is accounted for in the electron dynamics can   a laser pulse stop an electron in a head-on collision.

\subsection{Caldirola Equation}
Another proposal of considerable elegance  to generalize   LAD equations of motion was formulated  by Caldirola~\cite{Cald:79,Cald:78}
\beqn\label{Caldirola}
\mp\frac{m_e}{\delta t}\left[u^\alpha_\mp +u^\alpha \frac{u\cdot u_\mp}{c^2}\right]=
   \frac{e}{c}F^{\alpha\beta}u_\beta, 
\quad u_{\mp}\equiv u(t \mp\delta t), 
\quad \delta t\equiv \frac{4}{3}\frac{e^2}{m_ec^3}
\eeqn
Here the choice of sign determines if the electron is radiating (-) or absorbing (+) energy from the environment. Thus the LAD difficulty of run-away solutions is resolved by choosing the non-local form with upper sign in \req{Caldirola}. Further, the non-local form also suggests that consistent Maxwell-Lorentz dynamics can be  achieved by connecting with a discrete space-time. However, just like with LL modification of LAD, an action for the dynamics described by  \req{Caldirola} has  not been discovered.

The dynamics generated by each equation of motion \req{LADeqn}, \req{LLeom}  and \req{Caldirola} differ already at acceleration well below critical.  As a consequence, we can expect the radiation signatures to be distinguishable and the form of dynamical equations to be testable experimentally.  However, these proposed resolutions of the inconsistency suffer from their   ad hoc  format. We are missing a physics principle that would create a compelling format of action and thus particle and field dynamics, and therefore it is to be expected that none of these efforts represents a definitive theoretical description of charged particle dynamics.

\subsection{Experiments in Classical Domain}
Experiments are possible in kinematic domains where particle dynamics are classical rather than quantum and the predictions of \req{LADeqn}, \req{LLeom} and \req{Caldirola} differ.  Seeing that the couple of ad hoc modifications noted here are different in this domain, we believe that {\it any} modified classical dynamics are accessible to experiment. 

\begin{figure}
\centerline{\includegraphics[width=0.9\textwidth]{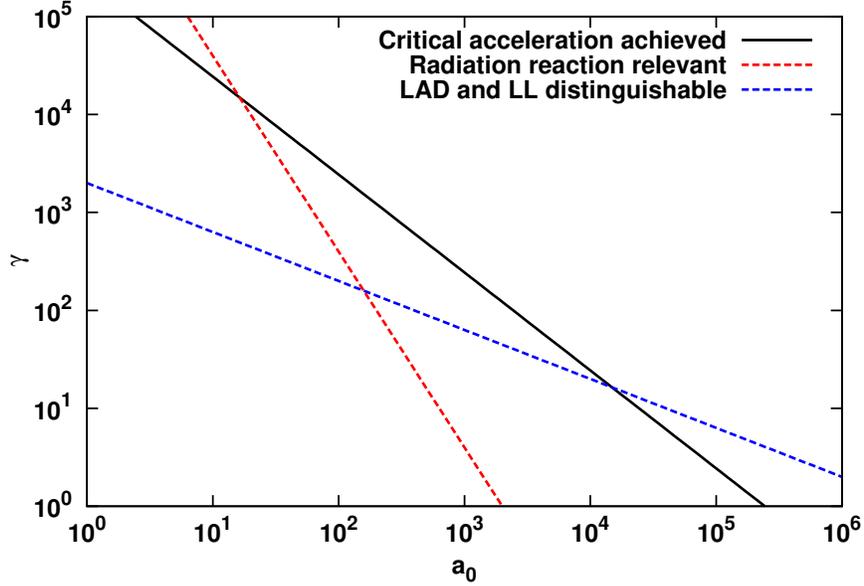}}
\caption{For the collision of a high energy $E/m_ec^2=\gamma $ electron with an intense laser pulse with normalized amplitude $a_0$ and frequency $\hbar\omega=4.8\:{\rm eV}$. Above the steeper dashed (red) line  radiation reaction is relevant \req{RRrelevant}; and above shallower dashed (blue) line LAD and Landau-Lifshitz equations of motion generate different dynamics.  Above the solid line (black), the electron experiences critical acceleration, however, quantum effects are expected to be important.\label{fig:rr}}
\end{figure}

For the case studied in detail~\cite{Hadad:2010mt} of a laser of amplitude characterized by the strength $a_0$ (in units of $mc^2$, see \req{azero}) colliding with a high energy electron $E=\gamma m_ec^2$, the condition for the relevance of radiation reaction is approximately given by
\beqn\label{RRrelevant}
a_0^2\gamma\gtrsim \frac{3}{2e^2}\frac{m_e}{\omega},
\eeqn
and LAD \req{LADeqn} and Landau-Lifshitz \req{LLeom} dynamics are distinguishable when
\beqn\label{LADLLdistinct}
a_0\gamma^2\gtrsim \frac{3}{2e^2}\frac{m_e}{\omega}.
\eeqn
On the other hand, the electron experiences critical acceleration when
\beqn\label{ecritaccel}
a_0\gamma=\frac{m_e}{\omega}
\eeqn
The different dependencies on electron and laser parameters ($\gamma$ and $a_0,\omega$, respectively) reveal the domain of interest, seen in Figure~\ref{fig:rr}.

As Figure~\ref{fig:rr} shows radiation reaction is relevant to classical dynamics, and there is a domain where radiation effects expected differ between forms of dynamical equations. Other observables such as pair production could be even more sensitive probes of radiation reaction classical dynamics. This means that the challenge to create a dynamical theory that consistently incorporates critical acceleration  must be already addressed within the classical physics domain. An important reason to retain focus on the classical domain is the evident connection to gravity. 

An important outcome of the effort to understand electromagnetism with critical acceleration is control of the abundant radiation produced by accelerated charges: naively one can say that a charge which is strongly  `kicked' loses for some time its electromagnetic mass component hidden in the field and must reestablish it. The purpose of the experiments is to understand how the true inertia and radiation define the electromagnetic part of mass. 

Nearly  100 years after the difficulty with electromagnetic theory was identified, experimental effort can   aid progress today~\cite{Labun:2010wf}.

\section{Mach, Inertia and the Quantum Vacuum}
Given that the EM equations of motion are generated from \req{EMaction} independently for particle dynamics and for field dynamics, we believe that in order to formulate  self-consistent particle-plus-field dynamics a connection of  two seemingly separate dynamical elements needs to be found. General relativity accomplishes this goal by removing acceleration altogether; particles are always in free fall, and  the geometry of the manifold is the field. 

The difference between the nature of classical gravity, described employing  general relativity theory, and the electromagnetic forces, provokes the question: {\it If a charged particle in a ``free-fall'' orbit bound by the gravitational field to travel around the Earth will not radiate, why should it emit synchrotron radiation  if the same orbit is established by other e.g. electromagnetic forces?}  This is a paradox of classical field theory arising from inherently different treatment of acceleration under gravitational and electromagnetic forces.

In view of how gravitational theory handles force, one approach to creating a consistent framework is to geometrize the electromagnetic (and in fact all gauge) interactions. This program puts all `forces' on the same footing as gravity, doing away with acceleration in a higher dimensional Kaluza-Klein space-time.  In this formulation, accelerated motion in our space-time may be a consequence of constraints imposed to reveal dynamics by projection onto the 3+1 dimensional hyper-surface. Despite many years of research this program has not been implemented to the point of practical applications, and there is no evidence that it will succeed.  

Moreover, geometrization of  EM interactions, similarly to gravity, does not resolve the  question  how to recognize accelerated versus non-accelerated observers in absence of Newton's absolute space. Implicit in the gravitational orbit  definition (see question above) is the presence of an (inertial) observer at asymptotically flat spatial infinity.  The authors are not aware of an electromagnetic theory that addresses the need to relate acceleration to an inertial Machian reference  frame. On the other hand, Mach,  Einstein~\cite{EinsteinMachltr} and many others have recognized the need to refer to a class of inertial frames of reference in order to agree on how to  measure  acceleration.  Mach's original proposal introduced the rest frame of the total mass of the universe, and an  analogous universal reference frame is the rest frame of the cosmic microwave background. 

Moving on from classical EM theory to quantum electrodynamics (QED) we observe a subtle difference even though the QED action is built on the same principles (gauge and Lorentz invariance) as the classical action \req{EMaction}, and the quantum action is organized to reduce to corresponding classical action.  As   seen e.g. in canonical quantization, any  quantum field theory requires the introduction  of a quantum vacuum state.  This vacuum state is a natural candidate for the inertial reference frame. 

The introduction of quantum theory implicitly   embraces reference to an inertial frame, a major conceptual advance compared to the classical theory.  Unfortunately, the text-book formulation of the classical limit of the quantum theory loses  the information  regarding the vacuum state and hence the  relationship to an inertial reference frame. We draw attention to the relativistic  Wigner function approach to classical limit,~\cite{BialynickiBirula:1991tx,Rafelski:1993uh}  which addresses this important challenge of retaining the information about the vacuum state in the classical  limit.  That was the good news, the bad news is that the dynamics of particle motion is connected closely with various possible processes of particle production, for which reason it has been exceedingly hard to make the relativistic Wigner method a practical tool. 

It is our belief  that the problems addressed here could be  partially or even completely resolved  by inclusion of the vacuum as an active ingredient in the classical dynamics of charged particles. This agrees well with the   vacuum structure defining the nature of the laws of physics; our understanding  of the influence of the vacuum has grown considerably in past 75 years. The Casimir force made vacuum fluctuations popular, and  spontaneous symmetry breaking defining the structure of electro-weak interactions in the Standard Model showed that interactions yesterday deemed `fundamental' can be effective.  Quark confinement (non-propagation in vacuum of color charge) defines the nature of  most of the mass of visible  matter in the Universe and arises from non-perturbative properties of the quantum vacuum.

Vacuum properties already play an integral role in electromagnetic theory. Below critical acceleration, the virtual possibility of pair creation leads to well-known vacuum polarization phenomena~\cite{Uehling:1935po,Euler:1935zz,Heisenberg:1935qt}, giving the quantum vacuum dielectric properties.  In Feynman diagram language, the real photon is decomposed into a bare photon and a photon turning into a virtual pair. The result is a renormalized electron charge smaller than the bare charge and a slightly stronger Coulomb interaction ($0.4\%$ effect at highest accessible experimental scales). 

The quantum vacuum state can be probed by experiments at critical acceleration, which in QED corresponds to the critical field strength \req{Ec}.  An electric field at the critical strength is expected to decay rapidly into particle-anti-particle pairs.  The decay process is non-perturbative, involving a large number of quanta ($N\hbar\omega \to \infty$), but has no classical analog or obvious limit.  This possibility of electromagnetic field decay gives rise to a temperature parameter analogous to but numerically different from the Unruh temperature detected by an observer accelerated through the vacuum~\cite{Labun:LeCosPA}.

\section{Conclusions}
We have discussed here the new opportunities to study foundational physics involving acceleration and described potential avenues to search for an extension of physics to understand critical acceleration phenomena. Experiments at and much beyond the critical acceleration \req{ac} can be today performed  in electron-laser pulse collisions. This in particular includes the `stopping' of relativistic charged particles by ultra-intense laser pulses. Such experiments  should help resolve the radiation reaction riddle in electromagnetic theory. We can expect a rich field of applications and theoretical insights to follow. 

The study of physics phenomena beyond critical acceleration should lead to a  better understanding of  the relation between General Relativity, Electromagnetism and Quantum Physics, and specifically to understanding of the relation of classical dynamics and inertia to the `Machian' inertial frame, the quantum vacuum.  The (effective) quantum field theory of known interactions, the Standard Model (SM), has incorporated the idea that a theory that describes forces must allow for a universal  inertial reference frame. However, this incorporation of an inertial  reference frame  is lost in the classical particle-motion theory that emerges in the naive classical limit.  We have shown that such a  classical theory does not provide a consistent framework to address supercritical forces/acceleration.

The inclusion of the quantum vacuum structure into the understanding of laws of physics  underpins our interpretation of the SM of all microscopic interactions. Since the quantum vacuum state  plays a pivotal role and  the SM requires input of tens of parameters,  it is the universal belief that the current   understanding of fundamental forces is  an effective theory.  This point of view agrees with our argument   that the present framework of classical forces in physics does not adequately describe particle dynamics at critical acceleration as it is not an adequate limit of the effective quantum theory from which supposedly it arises.  Both these observations could mean that a more fundamental theory has to be discovered before we can reconcile both SM complexities and classical theory deficiencies with the elegant Universe. The next step leading to this discovery, we suggest, is to study the critical acceleration by the way of relativistic electron-laser pulse collisions.

\section*{Acknowledgments} This work was supported by a grant from the  U.S. Department of Energy, DE-FG02-04ER41318.

\bibliographystyle{ws-procs975x65}

\bibliography{ws-pro-sample}

\end{document}